\documentclass[twocolumn,showpacs,amsmath,amssymb]{revtex4}
\usepackage{graphicx}
\usepackage{bm}
\begin{document}
\title{Inhomogeneous LOFF phase revisited for surface superconductivity}
\author{Victor Barzykin} 
\author{Lev P. Gor'kov}
  \altaffiliation[Also at ]{L.D. Landau Institute for Theoretical Physics,
Chernogolovka, 142432, Russia}
\affiliation{National High Magnetic Field Laboratory, 
Florida State University,
1800 E. Paul Dirac Dr., Tallahassee, Florida 32310 }
\begin{abstract}
We consider 2D surface superconductivity in high magnetic fields 
parallel to the surface. We demonstrate that the spin-orbit interaction at the 
surface changes the properties of the inhomogeneous superconducting
Larkin-Ovchinnikov-Fulde-Ferrell state that develops above fields given by 
the paramagnetic criterion. Strong spin-orbit interaction significantly 
{\it broadens} the range of existence of the LOFF phase, which takes 
the form of periodic superconducting stripes running along the field direction 
on the surface, leading to the anisotropy of its properties.
In connection with experiments by J.H. Sch\"{o}n {\em et al.}
[Nature \textbf{914}, 434 (2001)] on superconductivity of 
electrically doped films of the cuprate material CaCuO$_2$, we also discuss 
this problem for the d-wave pairing to indicate the possibility of a re-orientation
transition as the magnetic field direction is rotated in the plane parallel 
to the surface. Our results provide a tool for studying surface superconductivity
as a function of doping.
\end{abstract}
\vspace{0.15cm}

\maketitle

There has been renewed experimental and theoretical interest in the 
properties of
metallic states localized at a surface, and surface superconductivity (SSC). 
Surface states (a.k.a. Tamm's levels) are well-known from the physics of 
semiconductors. Numerous ARPES\cite{LaShell,Rotenberg}  data and 
STM studies of Friedel oscillations\cite{Petersen}
have now proven the existence of 2D metallic bands even at surfaces of metals. 
The bands are well-separated from the bulk and possess clear-cut 2D Fermi
surfaces. For heavy enough elements such Fermi surfaces 
are split further by strong spin-orbit (SO) interactions. For example, SO energy for 
electrons at the Fermi level for Au is estimated at 0.1 eV \cite{LaShell}, while for 
Li-doped surfaces of Mo and W its value increases up to 0.13 eV and 0.5 eV,
correspondingly \cite{Rotenberg}.

Islands of a surface superconducting phase were also observed for the 
surface-doped tungsten bronzes, WO$_3$:Na, at $T_c = 91.5 K$\cite{Reich} . The ARPES 
results mentioned above suggest that SSC may actually be a rather widespread
phenomenon.

It is noteworthy that the bulk WO$_3$ is an insulator at low doping\cite{Goodenough}. Recently 
a remarkable breakthrough has been accomplished by doping 
insulator surfaces electrically in the so-called Field-Effect Transistor (FET) geometry\cite{Schon}.
SSC with high T$_c$ was induced by both electronic and hole doping of films of 
the prototype cuprate material, CaCuO$_2$\cite{Schon1}. 

All this makes us believe that the search for SSC emerges as a new
and important development in studies of the properties of surfaces,
especially their metallic properties. As for the SSC itself, its
mechanisms are unknown, and may have nothing to do with the ones
in the bulk. Of a special challenge is the possibility of superconductivity
at surfaces of ordinary metals such as Cu \cite{Petersen} 
 or lithium-doped Mo and W\cite{Rotenberg}, and the 
influence of substrate  
on superconductivity in thin films.
Thus, it is necessary to return to a
more careful investigation of the low temperature properties
of surfaces with absorbed atoms and the role of adsorbed atoms
as dopants of carriers into the metallic surface zone 
(e.g., see Ref.\cite{Rotenberg}) like it was discovered for
WO$_3$:Na \cite{Reich}.

The major challenge to such a program lies in discerning
SSC and studying its properties. Even for insulators where doping by FET,
ideally, provides an effective control on surface properties,
experimental tools to probe SSC are limited in numbers. The Meissner effect 
due to surface superconducting islands would probably never produce a bulk 
screening. Thermodynamical probes are also difficult because of the smallness of
contributions from surface layers of atomic thicknesses. 
So far, e.g., in Ref. \cite{Schon},
surface superconductivity has only been detected by measuring resistivity dependence
on temperature in the perpendicular-to-plane magnetic fields.

In what follows we focus on destroying SSC by magnetic fields applied 
{\it parallel} to the surface. In high enough fields one should expect the 
appearance of a 2D version of inhomogeneous superconducting state known as 
Larkin-Ovchinnikov-Fulde-Ferrell phase (LOFF) \cite{LOFF}. 
It was also shown that two-dimensionality broadens the region of the LOFF 
state on the B-T phase diagram\cite{Bulaevskii}. Motivated by experimental findings mentioned above,
we investigate the peculiarities introduced into this phenomenon by the SO 
effect or non-s-wave pairing.
Experiments by Sch\"on et al. not only have proved FET to be an effective doping process.
Different levels of doping result in different $T_c$-s. Our results for SSC in parallel
magnetic fields are expressed in terms of this $T_c$, providing the tool for comparing
theoretical predictions with experiments by controlling the doping level.
Theoretically, there is no long range order in a 2D
superconductor. However, correlations are destroyed on the
exponentially large spatial scale, 
$R \sim \xi_0 exp(E_F/T)$, which would exceed the size of the film.

We employ below the weak-coupling BCS-like scheme by assuming that electrons 
interact via a weak short-ranged interaction, $U(\bm{r},\bm{r}')$. Then 
$T_c \ll \epsilon_F$, and only a narrow vicinity of the Fermi surface is 
involved. Thus, the interaction in the momentum representation can be 
taken in the form:
\begin{equation}
U(\bm{p},\bm{p}') = \sum_l U_l \chi_l(\bm{p})\chi_l(\bm{p'}),
\label{coupl}
\end{equation}
where $\bm{p}$, $\bm{p'}$ lie on the Fermi surface; the angular dependence 
is expressed through a complete set of basis functions 
$\chi_l(\bm{p})$ (index $l$ enumerates different representations, as in 
expansions over the spherical functions in a 3D isotropic model). 
Superconducting order parameter, the "gap", 
$\hat{\Delta}_{\alpha \beta} (\bm{p})$ is defined by the equation:
\begin{equation}
\hat{\Delta}_{\alpha \beta} (\bm{p}) = |U_l| \chi_l(\bm{p})
\int { d^3 p' \over (2 \pi)^3 } \chi_l(\bm{p'}) \left(T \sum_{\omega_n}
F_{\alpha \beta}(\bm{p'}; i \omega_n) \right),
\label{gap}
\end{equation}
where $F_{\alpha \beta}(\bm{p'}; i \omega_n)$ stands for the Fourier 
component of Gor'kov anomalous function:
\begin{equation}
\hat{F}_{\alpha \beta}(\bm{r}-\bm{r}', \tau - \tau')
= - \langle T_{\tau} \left(\hat{\Psi}_{\alpha}(\bm{r},\tau)
\hat{\Psi}_{\beta}(\bm{r}',\tau') \right)\rangle,
\label{F}
\end{equation}
and $U_l < 0$ is a constant in Eq.(\ref{coupl}) corresponding to the 
selected pairing channel. When the field operators for electrons are
re-written in momentum space,
$\hat{\Psi}_{\alpha}(\bm{r},\tau) = \sum_{\bm{p}} 
\hat{\Psi}_{\alpha}(\bm{p},\tau) e^{i \bm{p} \cdot \bm{r}},$
Eq.(\ref{gap}) becomes:
\begin{equation}
\hat{\Delta}_{\alpha \beta} (\bm{p}) = |U_l| \chi_l(\bm{p})
\int { d^3 p' \over (2 \pi)^3 } \chi_l(\bm{p'}) \langle
\hat{\Psi}_{\alpha}(\bm{p}',\tau)
\hat{\Psi}_{\beta}(-\bm{p}',\tau) \rangle.
\label{gap1}
\end{equation}
The two operators inside brackets in Eq.(\ref{gap1}) anti-commute. 
In the presence of the center of inversion (CI) the behavior of
$\chi_l(\bm{p'})$ at $\bm{p'} \longrightarrow - \bm{p'}$  alone
determines the symmetry of $\hat{\Delta}_{\alpha \beta} (\bm{p})$
(even (singlet) vs odd (triplet) parity pairing). For an s- or d-pairing 
the order parameter below $T_c$ has the form:
\begin{equation}
\hat{\Delta}_{\alpha \beta} (\bm{p}; \bm{q}) =
\Delta (\bm{q}, T) (i \sigma_y)_{\alpha \beta} \chi_l(\bm{p}),
\label{gap2}
\end{equation}
where the momentum $\bm{q}$ stands for the spatial dependence of the gap
amplitude, $\Delta (\bm{q}, T)$.

Surface always breaks the CI symmetry due to the difference between the 
"top" and the "bottom". The direction bulk-to-surface determines $\bm{n}$, a 
unit vector normal to the surface. {\it Qualitative} changes in the surface
electronic spectrum come about from the well-known Rashba term\cite{R60}:
\begin{equation}
\hat{h}_{SO} = \alpha (\bm{\sigma} \times \bm{ p} \cdot \bm{n}),
\label{rashba}
\end{equation}
which specifies SO interactions at a surface. Eq.(\ref{rashba}) lifts
the two-fold spin degeneracy for band electrons. The electron spectrum
now consists of two branches with two Fermi surfaces:
\begin{equation}
\epsilon_{\pm}(\bm{p}) = v_F(p-p_{F\pm}); 
\ \ p_{F\pm} = p_F (1 \pm {\alpha \over v_F})
\label{FS}
\end{equation}
Even though SO splitting, 
$2 \alpha p_F$, may be on a scale of tenths of eV \cite{LaShell,Rotenberg}, we assume 
that $2 \alpha p_F \ll \epsilon_F$. Below we speak of a strong or weak SO
meaning the relative values of $2 \alpha p_F$ and T$_c$.
\begin{figure}
\includegraphics[width=3in]{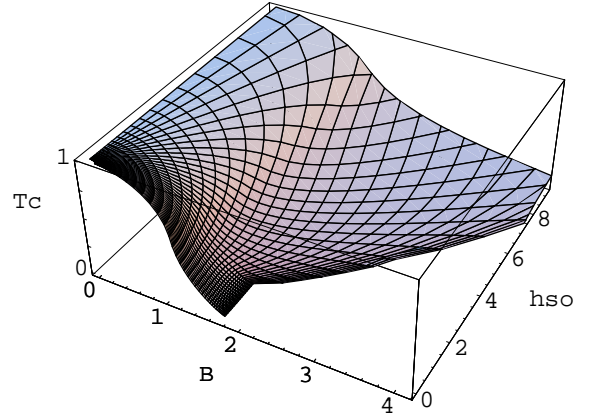}
\caption{Calculated $T_c$ as a function of magnetic field and 
spin-orbit interaction $\alpha p_F$ for a 2D 
surface superconductor. All values are in units of $T_{c0}$, 
$T_c$ for the 2D superconductor in the absence of magnetic field.
Thus, $B \rightarrow \mu_B B/T_{c0}$, 
$h_{SO} \rightarrow \alpha p_F/T_{c0}$, and $T_c \rightarrow T_c/T_{c0}$.}
\end{figure}

Some theory issues regarding superconductivity without CI due to the presence of
the SO term Eq.(\ref{rashba}) were first considered in Ref.\cite{Edelstein}
and more recently in Ref. \cite{GR}. The Gor'kov function
(in the momentum representation),
$F_{\alpha \beta}(\bm{p}, +0) = - \langle 
\hat{\Psi}_{\alpha}(\bm{p}) \hat{\Psi}_{\beta}(-\bm{p}) \rangle$,
that stands under integral in Eq.(\ref{gap1}) represents the
wave function for Cooper pairs in the condensate. In the presence of CI 
symmetry the latter can be classified according to the parity:
\begin{equation}
F_{\alpha \beta}(\bm{p};+0)=\left\{ \begin{array}{c}
i (\sigma_y)_{\alpha \beta} f(\bm{p}); \ \ (f(\bm{p}) \ even) \\
i [(\bm{d}(\bm{p}) \cdot \hat{\bm{\sigma}}) \hat{\sigma}_y ]_{\alpha \beta};
 \ \ (\bm{d}(\bm{p}) \ odd) \end{array} \right.
\label{triplet}
\end{equation}
With CI broken by non-zero SO term Eq.(\ref{rashba}), the pairing wave 
function becomes a mixture of even and odd terms. It is important to realize 
that while this mixing changes physical properties of the SC phase,
the gap order parameter, $\hat{\Delta}_{\alpha \beta}(\bm{p}; \bm{q})$,
preserves its singlet form Eq.(\ref{gap2}). For instance, s-pairing indeed induces 
the non-zero triplet component  Eq.(\ref{triplet}), as shown in 
Refs.\cite{Edelstein,GR}. However, the latter does not automatically generate a "triplet" gap, 
$\hat{\Delta}^t_{\alpha \beta}(\bm{p}; \bm{q})$. Indeed, re-writing the integration 
over $\bm{p}'$ in Eq.(\ref{gap1}) as 
$d^3 p' \rightarrow dS'_F d\xi$, where $\xi = v_F(p-p_F)$, we notice that the  triplet
$F$-component is odd in particle-hole transformation, $\xi \rightarrow - \xi$, and, hence,
the integrals of the form Eq.(\ref{gap1}) would only give small terms
of order $(\alpha p_F/\epsilon_F) \ll 1$. In other words, while SO interaction
may significantly change spin structure of the normal and anomalous Green 
functions,  the "gap", $\hat{\Delta}_{\alpha \beta}(\bm{p}, \bm{q})$, 
Eq.(\ref{gap}) preserves its usual form Eq.(\ref{gap2}) with 
$\chi_l(\bm{p}) = const$ for isotropic pairing, and 
$\chi_l(\bm{p}) \propto (p_x^2-p_y^2)$ for the d-wave pairing.

We have calculated $T_c$ numerically for the 2D superconductor 
as a function of magnetic field and spin-orbit interaction. The 
result is shown in Fig.1 for the magnetic field strictly parallel 
to the surface (to exclude diamagnetic currents)\cite{Bulaevskii}. 

A 1$^{st}$ order phase transition was initially expected between 
superconducting and normal states, defined by comparing their free
energies: $F_s(T) = F_n - \chi_N {B^2 \over 2}$,
which would determine the so-called paramagnetic critical field, $H_{par}$
\cite{Clog} ($\chi_N$ - the spin susceptibility in the normal
phase). The transition from normal to superconducting state is actually
(at lower temperatures) a second order transition into the LOFF state.
The details of the phase diagram in the vicinity of the $H_{par}$
were studied numerically in Ref.\cite{br}.
The LOFF phase boundary is determined by Eq.(\ref{gap}) or Eq.(\ref{gap1}),
linearized in $\Delta_{\alpha \beta}(\bm{q})$ at an extremal $\bm{q}$.
The corresponding expression for anomalous function linear in 
$\Delta_{\alpha \beta}(\bm{q})$ is obtained by solving the proper Gor'kov
equations:
\begin{equation}
F_{\alpha \beta}(\bm{p},\bm{q}; i\omega_n) =
- \hat{G}^{(0)}_{\alpha \nu}(\bm{p}; i\omega_n)
\hat{\Delta}_{\rho \nu}(\bm{p}, \bm{q}) 
\hat{G}^{(0)}_{\beta \rho}(-\bm{p}+\bm{q}; -i\omega_n),
\label{12}
\end{equation}
where $\hat{G}^{(0)}_{\alpha \beta}(\bm{p}; i\omega_n)$ is the normal
state Green function at non-zero $\hat{h}_{SO}(\bm{p})$ of Eq.(\ref{rashba}),
together with the Zeeman term, $\mu_B (\bm{\sigma} \cdot \bm{B})$:
\begin{equation}
[i \omega_n - \xi - \hat{h}_{SO}(\bm{p}) - \mu_B \hat{\bm{\sigma}}\bm{B}]
\hat{G}^{(0)}_{\alpha \beta}(\bm{p}; i\omega_n) = \hat{1}.
\end{equation}
The spin Hamiltionian on the left side, $\hat{H}=\hat{h}_{SO}(\bm{p}) + 
\mu_B \hat{\bm{\sigma}}\bm{B}$, may be easily diagonalized:
\begin{equation}
\tilde{\epsilon}_{\lambda}(p) - \xi = - \lambda
\sqrt{\alpha^2 p_F^2 + 2 \alpha p_{Fy}  \mu_B B + (\mu_B B)^2} \equiv
- \lambda \tilde{\epsilon}(p)
\label{spectrum}
\end{equation}
($\lambda = \pm 1$ for the two branches, $p_{Fy}$ is the y-axis projection of the Fermi momentum, 
and $B \parallel x$) with the 
eigenfunctions, spinors $\eta^{\lambda}(\bm{p})$ of the form:
\begin{equation}
\eta^{\lambda}(\bm{p}) = {1 \over \sqrt{2}} 
\left\{ \begin{array}{c}
1 \\ {\mu_B B - i e^{i \varphi(\bm{p})} p_F \alpha \over 
\lambda \tilde{\epsilon}(p)} \end{array} \right\}.
\label{spinor}
\end{equation}
Substitution of Eq.(\ref{12}) into Eq.(\ref{gap}) making use of 
Eq.(\ref{gap2}) results in a rather cumbersome expression which 
generalizes the corresponding  
Eq.(7) of Ref.\cite{Bulaevskii}. We sketch, therefore, 
only a few results for the low-T part of the phase diagram in Fig.1.
Below we discuss the main changes in the shape of the phase diagram, as 
introduced by SO coupling or anisotropy. As for the structure 
of the LOFF phase itself, we assume that the numerical analysis done in 
Ref.\cite{br} remains applicable, i.e. the order parameter in the LOFF state 
has the structure of periodic stripes.

We state in more detail our results for the limiting cases of strong
and weak SO interaction, which significantly simplify all calculations.

\bigskip
\noindent
{\it a) Strong SO: $\alpha p_F \gg \Delta(0)$}; 
subsequent analysis, which we do not provide here, leads after 
short calculations to our final results:
\begin{eqnarray}
\bm{q} \perp \bm{B}, & \ \ & | \bm{q}|= {2 \mu_B H_{c2} \over v_F}; 
\nonumber \\
\mu_B H_{c2} &=&  \sqrt{2 \Delta(0) \alpha p_F} \equiv 
\sqrt{{2 \pi \over \gamma} T_{c0} \alpha p_F}
\label{res}
\end{eqnarray}
One sees that SO interaction not only enhances the value of $H_{c2}$ in
comparison with the LOFF for 2D model of Ref.\cite{Bulaevskii} ,
but it also fixes the direction of the structure vector. The 
resulting stripe structure is parallel to the magnetic field direction and
has the perpendicular space periodicity 
$L= \sqrt{\pi \gamma \over 2 T_{c0} \alpha p_F} v_F$. 

According to \cite{GR}, in the limit of strong SO interaction the
spin susceptibility for parallel fields in superconducting state, 
$\chi_S(T)$ is non-zero and equal to ${1 \over 2} \chi_N$. This 
increases the critical paramagnetic 
field \cite{Clog} only by a factor of $\sqrt{2}$,
$\mu_B H_{par} = \Delta(0)$.
Comparing this with Eq.(\ref{res}), one sees that strong SO
significantly increases the area occupied by the LOFF state by
a factor of $\sqrt{\alpha p_F/T_{c0}} \gg 1$. Strong 
SO scattering by defects also enhanses $H_{c2}$\cite{Klemm}, 
but the LOFF state does not exist in presence of disorder.
Analytical expressions can also be obtained at low temperatures.
For the dependence of the transition temperature on magnetic field, $T_c(B)$, one 
obtains, at $T \ll T_{c0} \sqrt{T_{c0} /( \alpha p_F)}$:
\begin{equation}
T_c(B) = 5.784 \left(\sqrt{2 \pi T_{c0} \alpha p_F/\gamma} - \mu_B B\right)^3/(\alpha p_F)^2, 
\label{lowTstrongSO}
\end{equation}
and
\begin{equation}
T_c(B) =  {\pi T_{c0}^2 / (2 \gamma \mu_B B)}, \ \ T_{c0} 
\sqrt{T_{c0} / (\alpha p_F)} \ll T \ll T_{c0} 
\label{lowTstrongSO1}
\end{equation}
The expression for $T_c(B)$ 
also simplifies for small magnetic
fields near $T_{c0}$:
\begin{equation}
{T_{c0} - T_c \over T_{c0}} = {7 \zeta(3) \over 8 \pi^2} 
{(\mu_B B)^2 \over T_{c0}^2}
\label{weakf}
\end{equation}
Note that while the spin-orbit interaction splits the Fermi surface, quasiparticles
with the same band spin index Eq.(\ref{spectrum}) form Cooper pairs in the 
superconducting state, so that spin-orbit interaction alone does not change $T_c$.

\bigskip
\noindent
{\it b) Weak SO: $\alpha p_F \ll \Delta(0)$}; unlike in case of strong SO,
the Cooper pair is formed mainly by pairing of electrons from the FS's 
with different spin indices. The LOFF phase in a 2D superconductor with no
SO interaction was first analysed in Refs \cite{Bulaevskii,br}. According
to Ref.\cite{Bulaevskii}, $\mu_B H_{c2} = \Delta(0) = \sqrt{2} H_{par}$,
$v_F q = 2 \mu_B H_{c2}$.

This result is zero order in spin-orbit interaction.
The direction of $\bm{q}$ is {\it not fixed} with respect to $B$. Analysis
to the second order in $\alpha p_F$ results in an anisotropy term which 
again fixes the vector $\bm{q}$, as in strong SO case Eq.(\ref{res}),
perpendicular to
the direction of the magnetic field. Indeed, for the critical field as a 
function of the angle $\beta$  between $\bm{q}$ and $\bm{B}$ we find:
\begin{equation}
\mu_B H_{c2} = \Delta(0) - {\alpha^2 p_F^2 \over 2 \Delta(0)} cos^2\beta,
\end{equation}
i.e. the maximum for $H_{c2}$ is reached for $\beta = \pm {\pi \over 2}$.

For small magnetic fields near $T_{c}$ we get:
\begin{equation}
{T_{c0} - T_c \over T_{c0}} = {7 \zeta(3) \over 4 \pi^2}
{(\mu_B B)^2 \over T_{c0}^2}
\label{weakf1}
\end{equation}
A quadratic dependence on $B$ remains valid for fields $\mu_B B \sim \alpha p_F$.
However, note the factor two difference between
Eqs.(\ref{weakf}) and (\ref{weakf1}): suppression of $T_c$ by magnetic field 
turns out to be slower in case
of strong spin-orbit interaction than in case when the spin-orbit
interaction is weak. 

Although a more complicated LOFF periodic superstructure is possible,
the energy considerations of Ref.\cite{LOFF} have shown in 3D that the
stripe phase is energetically more favorable. A detailed study done
numerically in 2D \cite{br} has shown a more complicated than just a 
sinusoidal shape of the order parameter. We assume that the results of
Ref.\cite{br} remain valid in all cases considered above, so that
the LOFF state preserves its striped order parameter form. 

The major role of SO is in fixing of the LOFF superstructure. Anisotropy fixes
the orientation of the LOFF stripes as well (see below). A
1$^{st}$ order re-orientation transition may be expected corresponding to 
an abrupt change in the direction of the superconducting stripes 
(similar to a spin-flop transition), as the magnetic field is rotated in the 2D plane.

In the above discussion so far we have neglected any anisotropy at all. Meanwhile,
the anisotropy is, of course, important. We address the issue of
the effect of pinning stripe direction to the particular crystal axis 
only for the d-wave order parameter, since the latter is intrinsically 
anisotropic. Stripes may orient themselves along the directions of the gap 
maximums. It can be easily shown that $H_{c2}$ for the $d_{x^2-y^2}$ 
superconducting order parameter takes the form: 
$\mu_B H_{c2} = {\pi \over \gamma}  e^{1/4} T_{c0}$.
The critical field for the LOFF phase of d-wave is somewhat higher than 
for s-wave (without the SO term Eq.(7)), as 
supeconducting stripes get pinned to the crystal axes by the form of 
the order parameter and the direction of the magnetic field. Reorientation
transition/twinning is also expected for any other cause.

In summary, we have shown: i) that inhomogeneous state in parallel fields extends
considerably the low temperature phase diagram of surface superconductivity with 
increased spin-orbit interaction; ii) all SC characteristics of the phase diagram 
in the $(B,T)$ plane can be expressed in terms of $T_{c0}$, the critical temperature in 
the absence of the field, which is controlled by doping, and, hence, not only allows
a data comparison in the broad range of FET-doping, but may also serve as a method to 
extract the value of the SO interaction; iii) LOFF state properties are strongly anisotropic 
in the plane with respect to field direction - this, for example, can be seen by measuring 
the anisotropy of the AC susceptibility signal; iv) the indispensible feature of the LOFF state
must be the re-orientation transitions at the field rotation in the plane caused by locking of the
LOFF order parameter by anisotropy.

We would like to thank L. N. Bulaevskii for useful discussions and V. M. Edelstein
for a few references, including Ref.\cite{Edelstein}. 
This work was supported by NHMFL through the NSF Cooperative agreement 
No. DMR-9521035 and the State of Florida.


\begin{thebibliography}{}
\bibitem{LaShell} S. LaShell, B. A. McDougall, and E. Jensen, Phys. Rev. Lett.
\textbf{77}, 3419 (1996).
\bibitem{Rotenberg} E. Rotenberg, J. W. Chung, and S. D. Kevan, Phys. Rev. Lett.
\textbf{82}, 4066 (1999).
\bibitem{Petersen} L. Petersen {\em et al.}, 
Phys. Rev. B \textbf{57}, R6858 (1998).
\bibitem{Reich} S. Reich and Y. Tsabba, Eur. Phys. J. B \textbf{9}, 1 (1999);
Y. Levi {\em et al.}, Europhys. Lett. \textbf{51}, 564 (2000).
\bibitem{Goodenough} J.B. Goodenough, Progr. Solid State Chem. \textbf{5}, 149 
(1971).
\bibitem{Schon} J. H. Sch\"{o}n, Ch. Kloc, R. C. Haddon, and B. Battlog,
Science \textbf{288}, 656 (2000); J. H. Sch\"{o}n, Ch. Kloc, and B. Battlog,
Nature \textbf{406}, 702 (2000).
\bibitem{Schon1} J. H. Sch\"{o}n, M. Dorget, F.C. Beuran, X.Z. Zu, 
E. Arushavov, C. Deville Cavellin, and M. Lagn\"{e}s,
Nature \textbf{414}, 434 (2001).
\bibitem{LOFF} A. I. Larkin and Yu. N. Ovchinnikov, Sov. Phys. JETP \textbf{20}, 762
(1962); P. Fulde, R. A. Ferrell, Phys. Rev. \textbf{135}, A550 (1964).
\bibitem{Bulaevskii}  L. N. Bulaevskii, Sov. Phys. JETP \textbf{37}, 1133 (1973)
[ZhETF \textbf{64}, 2241 (1973)]; Sov. Phys. JETP \textbf{38}, 634 (1974)
[ZhETF \textbf{65}, 1278 (1973)]; M. Houzet, A. Buzdin, L. Bulaevskii, and M. Maley,
Phys. Rev. Lett. \textbf{88}, 227001 (2002).
\bibitem{R60} E. I. Rashba, Sov. Phys. - Solid State, \textbf{2}, 1109 (1960);
Yu. A. Bychkov and E. I. Rashba, Sov. Phys. - JETP Lett. \textbf{39}, 78 (1984).
\bibitem{Edelstein} V. M. Edelstein, Sov. Phys. JETP \textbf{68}, 1244 (1989).
\bibitem{GR} L.P. Gor'kov and E. I. Rashba,
Phys. Rev. Lett. \textbf{87}, 037004 (2001) (The problem has also been treated
by L.N. Bulaevskii, A.A. Guseinov, and A.I. Rusinov,
Sov. Phys. JETP \textbf{44}, 1243 (1976) [ZhETF \textbf{71}, 2356 (1976), although their results contain errors).
\bibitem{Clog} A. M. Clogston, Phys. Rev. Lett. \textbf{9}, 266 (1962);
B. S. Chandrasekhar, Appl. Phys. Lett. \textbf{1}, 7 (1962).
\bibitem{br} H. Burkhardt and D. Rainer, Ann. Physik \textbf{3}, 181 (1994). 
This work claims that, instead of the 1-st order transition predicted in
Ref.\protect{\cite{Clog}}, the LOFF state transforms \textit{continuously} 
into a uniform superconductor by developing wide regions of a constant 
order parameter separated by narrow domain walls where the order parameter 
changes sign.
\bibitem{Klemm} R. A. Klemm, A. Luther, and M. R. Beasley,
Phys. Rev. B \textbf{12}, 877 (1975). 


\end{thebibliography}
\end{document}